\documentclass{acm_proc_article-sp}
\usepackage{listings}
\usepackage{tabularx}
\usepackage{calc}
\usepackage{enumitem}
\usepackage{booktabs}
\usepackage[table]{xcolor}
\usepackage{rotating}

\begin{document}

\title{Performance comparison between Java and JNI for optimal implementation of computational micro-kernels}

\numberofauthors{3}
\author{
\alignauthor
Nassim Halli \\
\affaddr{Univ. Grenoble, France}\\
\affaddr{Aselta Nanographics, France}\\
\email{nassim.halli@aselta.com}
\alignauthor
Henri-Pierre Charles\\
\affaddr{CEA-LIST, France}\\
\alignauthor
Jean-Fran{\c c}ois Mehaut\\
\affaddr{Univ. Grenoble}\\
}
\maketitle
\begin{abstract}
General purpose CPUs used in high performance computing (HPC) support a vector instruction set and an out-of-order engine dedicated to increase the instruction level parallelism. Hence, related optimizations are currently critical to improve the performance of applications requiring numerical computation. Moreover, the use of a Java run-time environment such as the HotSpot Java Virtual Machine (JVM) in high performance computing is a promising alternative. It benefits from its programming flexibility, productivity and the performance is ensured by the Just-In-Time (JIT) compiler. Though, the JIT compiler suffers from two main drawbacks. First, the JIT is a black box for developers. We have no control over the generated code nor any feedback from its optimization phases like vectorization. Secondly, the time constraint narrows down the degree of optimization compared to static compilers like GCC or LLVM. So, it is compelling to use statically compiled code since it benefits from additional optimization reducing performance bottlenecks. Java enables to call native code from dynamic libraries through the Java Native Interface (JNI). Nevertheless, JNI methods are not inlined and require an additional cost to be invoked compared to Java ones. Therefore, to benefit from better static optimization, this call overhead must be leveraged by the amount of computation performed at each JNI invocation. In this paper we tackle this problem and we propose to do this analysis for a set of micro-kernels. Our goal is to select the most efficient implementation considering the amount of computation defined by the calling context. We also investigate the impact on performance of several different optimization schemes which are vectorization, out-of-order optimization, data alignment, method inlining and the use of native memory for JNI methods.
\end{abstract}


\section{Introduction and Motivation}

The initial motivation for this study is to improve the performance of a Java application which intensively uses a small set of computational micro-kernels. Java has become an important general-purpose programming language and industry expresses great interest in using it for scientific and engineering computation \cite{ref_javaHpc2, ref_javaHpc}. High performance scientific Java, however, suffers from the lack of optimized libraries and low-level control \cite{ref_javaHpc2}. This is in part due to the machine-independent bytecode representation of Java which prevents from target specific optimization at the source code level. In a dynamically compiled environment, the optimizations are delegated to the JIT compiler which is in charge of translating bytecode into machine code at run-time, using dynamic information. JIT optimization like vectorization are essentials for performance and still remain a major concern \cite{ref_phase, ref_inliningHeuristics, ref_c2}. They enable programmers to maintain a machine-independent code and save them from writing several versions for different target architectures. That is why in Java applications pure Java methods are often preferred to native ones. \\
Though, the JIT compiler suffers from two main drawbacks. First, the JIT is a black box for developers. We have no control over the generated code nor any feedback from its optimization phases like vectorization. Secondly, the time constraint narrows down the degree of optimization \cite{ref_vapor, ref_phase} compared to static compiler like GCC or LLVM. As a result, it can be interesting to use statically compiled code to benefit from a deeper optimization for performance bottlenecks that usually represent a small part of the application. Java applications enable to call native code from dynamic libraries through the Java Native Interface (JNI). Nevertheless, JNI methods are not inlined and require an additional cost to be invoked compared to Java ones. Thus, to benefit from better static optimization, this call overhead must be leveraged by the amount of computation performed at each JNI invocation. This is what we call the flop-per-invocation \\
Considering these aspects (summarized in Table \ref{Drawbacks}) selecting the most efficient implementation between JNI and Java requires an advanced analysis that takes into account the flop-per-invocation. In this paper we tackle this problem and we propose to do this analysis for a set of micro-kernels in order to select the most efficient implementation for a given range of flop-per-invocation. We also investigate how the flop-per-invocation impacts the performance of several different optimization schemes which are vectorization, out-of-order optimization, data alignment, method inlining and the use of native memory for JNI methods. \\
The study is performed using the Java HotSpot Virtual Machine over a Sandy Bridge x86-64 architecture which supports the AVX (Advanced Vector Extensions) instruction set for vectorization. Furthermore, we use GCC for static compilation of the micro-kernels. Section \ref{Performance_metrics} provides a background about performance analysis and Section \ref{Code_optimization} provides a background about the optimization considered for this study. Section \ref{Benchmark_methodology} presents our benchmark methodology and experimental conditions. In Section \ref{Results} we expose and discuss the obtained results.

\begin{table}[h]
\centering
\scriptsize
\caption{\label{Drawbacks}Performance drawbacks and benefits overview}
\begin{tabular}{|m{1.1cm}| m{2cm}| m{2cm}|}
\cline{2-3}
\multicolumn{1}{m{1.1cm}|}{}& \textbf{Drawbacks} & \textbf{Benefits} \\
\hline
\textbf{JIT} & Lower level of optimization & No call overhead \\
\hline
\textbf{JNI} &  Higher call overhead  & Higher level of optimization \\
\hline
\end{tabular}
\end{table}

\section{Performance metrics}\label{Performance_metrics}

\subsection{Arithmetic intensity}
We call flop-per-invocation the number of floating point operations (flop) performed by the kernel at each invocation. The performance of a kernel implementation is measured in flop/s (i.e. the number of floating point operations performed per second). We have the following equation:
\begin{equation}
\label{ai_eq}
F = AI \times M
\end{equation}
Where $F$ is the flop-per-invocation, $M$ is the memory-per-invocation in byte corresponding to the amount of input and output memory processed by the kernel. $AI$ is the arithmetic intensity in flop/byte which is the ratio of the flop-per-invocation by the memory-per-invocation.\\
The arithmetic intensity allows to locate micro-kernel bottlenecks \cite{ref_roofline}. If the bottleneck is the memory bandwidth then the kernel is called memory-bound. Otherwise the bottleneck is the CPU throughput and the kernel is called CPU-bound. Indeed, we can write the following inequality:
\begin{equation}
\label{memory_bound}
\Pi > AI \times \beta
\end{equation}
Where $\Pi$ is the CPU peak performance in flop/s and $\beta$ is the peak memory bandwidth in byte/s. If Equation \ref{memory_bound} is satisfied then the kernel is memory-bound (otherwise CPU-bound). Since modern architectures use a multi-level cache hierarchy, the memory bandwidth depends on the data location over this hierarchy. This means that the memory bandwidth depends on the memory-per-invocation which impacts cache efficiency therefore increasing the memory-per-invocation leads to reducing the memory bandwidth.
\subsection{Performance profile}
\label{perf_profile}
We define the performance profile of a kernel implementation as the performance function of either the flop-per-invocation, or the memory-per-invocation (since there are linearly related). By plotting the performance profile of a kernel implementation for a wide range of memory-per-invocation, we can observe, for example, the memory bandwidth bottleneck.
\subsection{Theorical impact of the invocation cost}
\begin{figure}[h]
\centering
\includegraphics[width=\columnwidth]{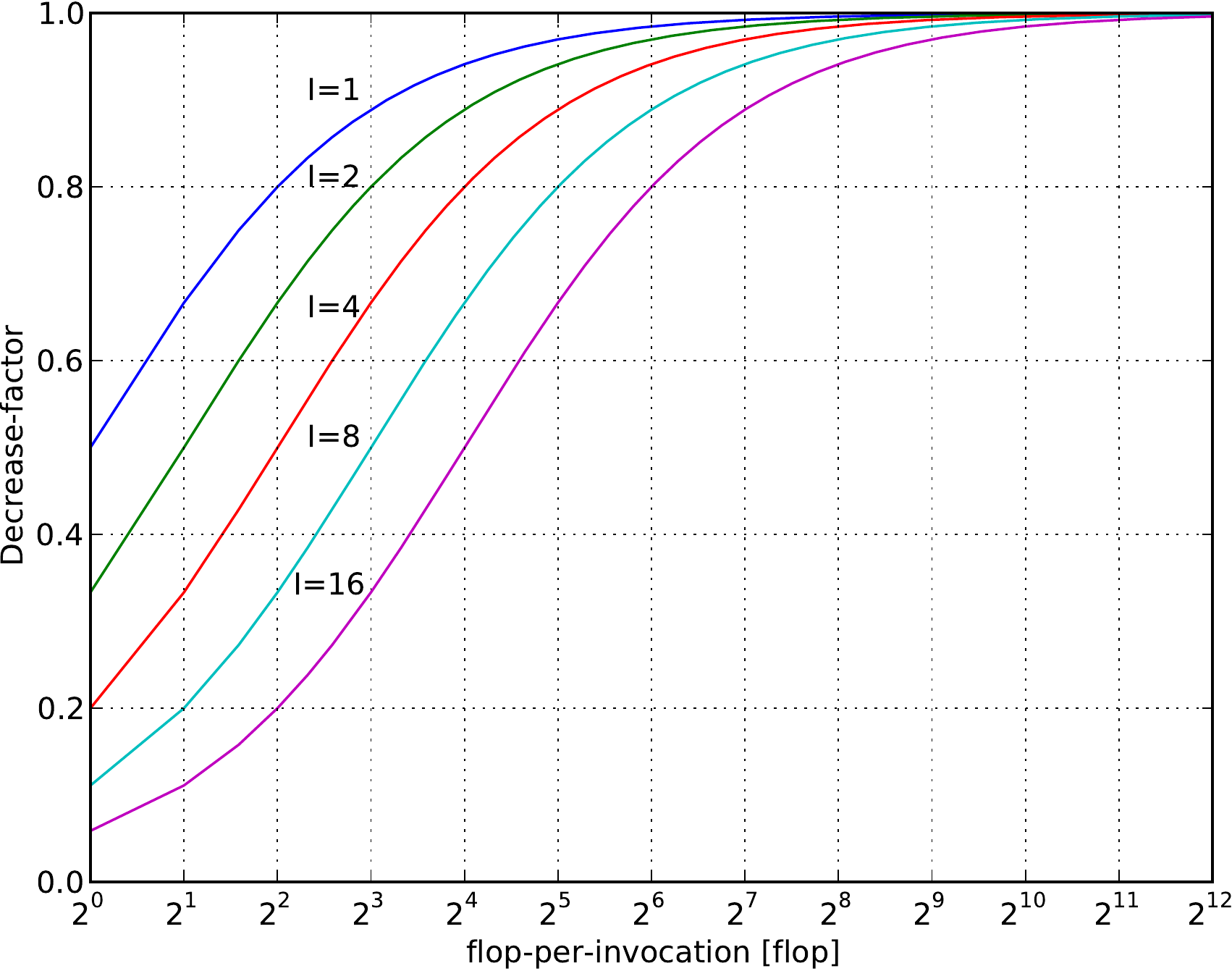}
\caption{Theoretic performance profile : decrease-factor function of the flop-per-invocation for different value of the invocation cost $I$}
\label{DF}
\end{figure}

A simple model to describe the invocation cost impact on performance when the flop-per-invocation decreases is given by the following equation:
\begin{equation}
P=\frac{P_{max} F}{I + F}
\end{equation}
Where $P$ is the performance, $F$ is the flop-per-invocation, $P_{max}$ is the peak performance reached by the kernel and $I$ is a parameter which describes the invocation cost in flop. We define the decrease-factor $DF$ as followed:
\begin{equation}
DF=\frac{P}{P_{max}}=\frac{F}{I + F}
\end{equation}
Figure \ref{DF} shows the decrease-factor as a function of the flop-per-invocation for different values of $I$. We can see that for relatively small flop-per-invocation, the invocation cost has a large impact on performance. Accordingly, we must consider both the flop-at-invocation and the JNI invocation cost to select the most efficient implementation.

\section{Code optimization}
\label{Code_optimization}
We consider two kinds of optimization, The first ones are asymptotic optimizations that are significant when the flop-per-invocation grows. In the second kind we have the optimizations that reduce the invocation cost of a method and which are significant for lower flop-per-invocation. They are described respectively in Section \ref{asympOpt} and \ref{invocOpt}.
\subsection{Asymptotic optimization}
\label{asympOpt}
\subsubsection{Vectorization and out-of-order optimization}
Since microprocessor performance is limited by a maximum clock frequency the trend goes towards increasing processor throughput by handling multiple data in parallel. The principle of vectorization is to pack data into a vector register then perform one operation over all the data. In this study we consider the AVX instruction set available in the 64-bit Intel Sandy Bridge architecture. AVX enables to use 256 bits registers called YMM. It allows to perform four double precision operations at one time. Additionally to vector units, the out-of-order (o-o-o) execution can increase the level of parallelism by dispatching independent instructions over parallel execution units available in different ports. For example the Sandy Bridge has six execution ports. Ports 0, 1 and 5 are for arithmetic and logic operations and support 256 bit vector operation. Ports 2 and 3 are two identical memory-load ports while port 4 is for memory-store. Increasing the level of parallelism with o-o-o is done by breaking long dependency chains. Hence, the o-o-o engine can evenly distribute independent operations between the different execution ports. Mixing o-o-o and vectorization leads to an optimal instruction level parallelism. Table \ref{opts} shows an example of source code optimization using vectorization and o-o-o for a horizontal sum kernel.

The HotSpot Virtual Machine run-time uses two compilers, the client compiler and the server compiler. The server version \cite{ref_c2}, which is the focus in this study, is a highly optimizing bytecode compiler which targets long running applications and aims to improve asymptotic performance. The server compiler supports auto-vectorization, it internally uses Super-word Level Parallelism (SLP) \cite{ref_SLP}. SLP detects groups of isomorphic instructions in basic blocks. This results in a lightweight vectorization approach suitable for Just-In-Time compilation. However, this approach cannot optimize across loop iterations and cannot vectorize loop-carried idioms like reduction. As a result, the JIT can not mix o-o-o and vectorization. Additionally the JIT compiler can be seen as a black box. Due to its portability specification Java does not provide machine-specific intrinsics like AVX intrinsics and of course, we can not inline assembly code directly in the Java source code. Additional meta-information, for example to ensure the compiler that pointer won't alias, cannot be passed to the JIT compiler. Finally, the JIT does not provide any feedback about its vectorization decisions which are invisible to programmers. Thus, to ensure that the code has been properly vectorized the generated code must be examined. Table \ref{Drawbacks2} summarizes the Java vectorization drawbacks.

\begin{table}[h]
\centering
\scriptsize
\caption{\label{Drawbacks2}Java vectorization drawbacks summary}
\begin{tabular}{|m{2cm}| m{2.5cm} | m{2.5cm}|}
\cline{2-3}
\multicolumn{1}{m{2cm}|}{} & \textbf{HotSpot server JIT} & \textbf{GCC}\\
\hline
\textbf{Pointer Aliasing} & Possibly unresolved & Resolved with run-time checks or restrict keyword \\
\hline
\textbf{Data alignment} &  Possibly unresolved & Alignment at allocation \\
\hline
\textbf{Reduction idiom}  &  Unvectorized at the current state & Vectorized \\
\hline
\textbf{Combine Out-of-orders and vectorization} &  Not supported & Supported \\
\hline
\textbf{Auto-vectorization Feedback} &  No feedback & Vectorization decisions and profitability threshold\\
\hline
\textbf{Source code vectorization} &  Not supported & AVX Intrinsics or inline assembly\\
\hline
\end{tabular}
\end{table}

\begin{table}[h]
\centering
\scriptsize
\caption{\label{opts}Core vectorization and out-of-order optimizations for a horizontal sum reduction kernel in double precision}
\begin{tabular}{|m{0.1cm}| m{2.4cm} | m{4.3cm}|}
\cline{2-3}
\multicolumn{1}{m{0.1cm}|}{}
&
\textbf{No vectorization}
&
\textbf{Vectorization (AVX intrinsics)}
\\ \hline
\begin{turn}{90}
\textbf{No o-o-o}
\end{turn}
&
\begin{lstlisting}[language=C, basicstyle=\tiny]
for(i=0;i<n;++i){
sum+=a[i];
}
\end{lstlisting}
&
\begin{lstlisting}[language=C, basicstyle=\tiny]
for(i=0;i<n;i+=4){
p=mm256_loadu_pd(&(a[i]));
sum=mm256_add_pd(sum,p);
}
\end{lstlisting}
\\ \hline
\begin{turn}{90}
\textbf{o-o-o}
\end{turn}
&
\begin{lstlisting}[language=C, basicstyle=\tiny]
for(i=0;i<n;i+=4){
sum0+=a[i];
sum1+=a[i+1];
sum2+=a[i+2];
sum3+=a[i+3];
}
\end{lstlisting}
&
\begin{lstlisting}[language=C, basicstyle=\tiny]
for(i=0;i<n;i+=16){
p0=mm256_loadu_pd(&(a[i]));
p1=mm256_loadu_pd(&(a[i+4]));
p2=mm256_loadu_pd(&(a[i+8]));
p3=mm256_loadu_pd(&(a[i+12]));
sum0=mm256_add_pd(sum0,p0);
sum1=mm256_add_pd(sum1,p1);
sum2=mm256_add_pd(sum2,p2);
sum3=mm256_add_pd(sum3,p3);
}
\end{lstlisting}
\\ \hline
\end{tabular}
\end{table}

\subsubsection{Data alignment}
\begin{figure}[h]
\centering
\includegraphics[width=\columnwidth]{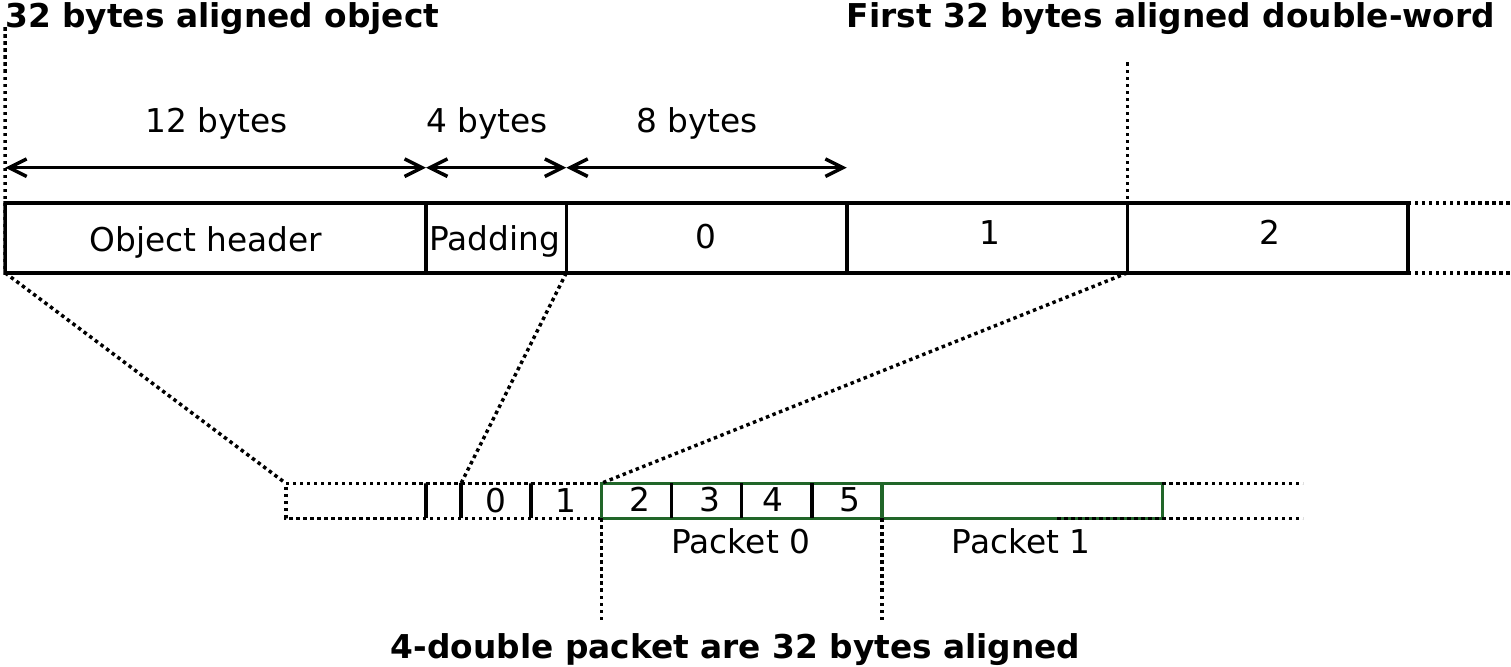} %
\caption{Java double array 32-byte aligned. The first 32-byte aligned packet begins at the offset 2}
\label{da_alignment}
\end{figure}

The proper alignment of packed data is required to observe vectorization benefit. In the considered case of a Sandy Bridge architecture, packets are 32-byte wide (which corresponds to 4 packed double). To ensure a benefit, packets must be aligned on 16 bytes which is the higher granularity for load and store instructions. With GCC we can use aligned AVX instructions if data packets are 32-byte aligned and pointers can be properly aligned using the \textit{posix\_memalign} system call. In Java we cannot explicitly align primitive arrays and the JIT is in charge of aligning memory access. To do so, compilers perform loop peeling i.e. they begin to load or store packed data at the first aligned offset. However in some cases when one loop iterates over several arrays which are offset relative to each other, peeling cannot resolve alignment for all the arrays. This causes a serious performance penalty which remains unpredictable. As we want to control the alignment of a Java object and its memory layout, we can set statically a base offset to ensure aligned memory access. Figure \ref{da_alignment} shows an example of a Java double array explicitly aligned on 32-byte by setting a base offset at 2 instead of 0.

\subsection{Optimizations that reduce the invocation cost}\label{invocOpt}
\subsubsection{Java method inlining}
Inlining is a core optimization especially in dynamically compiled environment. It allows to achieve significant optimization by specializing a method considering its calling context. Inlining also eliminates the invocation cost. In this study, we only consider inlining as an optimization that eliminates the invocation cost. To do so we consider kernels which show a constant complexity through inlining. Thereby the comparison between JNI and Java is also consistent because the experiment is independent of the calling context.
\subsubsection{JNI and native memory}
JNI is a  specification which allows applications to call statically compiled methods from dynamic libraries and to manipulate Java objects from there. JNI suffers from an additional invocation cost \cite{ref_GNFI, ref_JNIPerf, ref_inlineJNI}. Invoking a native target function requires two calls. The first is from the Java application to the JNI wrapper which sets up the native function parameters. The second is from the JNI wrapper to the native target function. But the most significant source of overhead occurs during the invocation of JNI callbacks. An example is to get an array pointer from the Java heap. A callback pays a performance penalty because two levels of indirection are used: one to obtain the appropriate function pointer inside a function table and one to invoke the function. Yet, these callbacks remains necessary to work with Java objects while avoiding memory copies that are even more expensive. In addition to the Java heap, where objects are allocated and automatically removed by the garbage collector, Java allows programmers to allocate chunks of memory out of the Java heap. This kind of memory is called native memory. We explore the use of native memory as an optimization technique in order to avoid JNI callbacks and so reduce its invocation cost.

\section{Benchmark methodology}\label{Benchmark_methodology}

For each kernel we provide several implementations that use the optimization techniques detailed in Section \ref{Code_optimization}. For each implementation, we measure and plot its performance profile (defined in Section \ref{perf_profile}). We aim to select the best implementation for a given range of flop-per-invocation and also to analyze the impact of a given optimization.

\subsection{Experimental Conditions}
Performance results are measured on an Intel(R) Sandy Bridge Core(TM) i5-2500 3.30GHz CPU. The Linux system version is 2.6.3 and we use the OpenJDK Runtime Environment 1.8.0, 64-Bit Server VM (build 25.0-b70, mixed mode). For JNI implementation we use GCC 4.4.7. The CPU has three levels of cache, for each we give its size and approximate latency:
\begin{itemize}
\item The L1 data cache is 32kB, its approximate latency is around 4 cycles.
\item The L2 cache is 256kB, its approximate latency is around 12 cycles
\item The L3 cache is 6MB, its approximate latency is around 30 cycles
\end{itemize}
The theoretical performance peak is 41.6 Gflop/s and we measure the approximate memory bandwidth peak around 40 GB/s. According to Equation \ref{memory_bound} in Section \ref{Performance_metrics} a kernel is memory-bound (resp. CPU-bound) if its arithmetic intensity is lower (resp. greater) than 1 flop/byte.

\subsection{Kernels sample and notations}
We select basic micro-kernels, described in Table \ref{kernels}, for our experimentation. Tested kernels computes in double precision. Vectorization is always profitable since data are already packed into memory. \\
For each kernel we provide several implementation. An implementation is described by the following label:
\begin{equation}
\label{label}
\nonumber
Type\_InvocationOpt\_AsymptoticOpts
\end{equation}
Where $Type$ is the type of the implementation, $InvocationOpt$ is the optimization that reduces the invocation cost and $AsymptoticOpts$ are the asymptotic optimizations. Table \ref{ImplDesc} details the values taken by each attributes.

\begin{table}[h]
\centering
\scriptsize
\caption{\label{ImplDesc}Implementation label description}
\begin{tabular}{|m{1.8cm}| m{2.7cm} | m{2.7cm}|}
\hline
$Type$ & $InvocationOpt$ &  $AsymptoticOpts$ \\
\hline
\begin{itemize}[leftmargin=*]
\item \textit{java}: pure Java methods + JIT
\item \textit{jni}: JNI methods + GCC
\end{itemize}
&
\begin{itemize}[leftmargin=*]
\item \textit{inline}: inlining for Java methods
\item \textit{native}: native memory for JNI methods
\item empty if no optimization
\end{itemize}
&
\begin{itemize}[leftmargin=*]
\item \textit{vect}: vectorization with aligned data
\item \textit{vect\_unalign}: vectorization with unaligned data
\item \textit{ooo}: out-of-order optimization
\item empty if no optimization
\end{itemize}
\\
\hline
\end{tabular}
\end{table}

\begin{table}[h]
\centering
\scriptsize
\caption{\label{kernels}Micro-kernels Sample}
\begin{tabular}{|m{3.5cm}| m{1.5cm}| m{2cm}|}
\hline
\textit{kernels description} & \textit{arithmetic intensity} & \textit{properties} \\
\hline
\textbf{Array addition}. Adds an array inside the other & 1/16 & memory-bound \\
\hline
\textbf{Horizontal sum}. Sums the values of an array & 1/8 & memory-bound, unvectorized by the JIT \\
\hline
\textbf{Horner coefficient-1st}. Computes a 64-degree polynomial on N data using the Horner method. Loops over the coefficients array at first. & $\frac{192N}{8(64+2N+1)}$ & cpu-bound ($N>4$)\\
\hline
\textbf{Horner data-1st}. Computes a 64-degree polynomial on N data using the Horner method. Loops over the data array at first. & $\frac{192N}{8(64+2N+1)}$ & cpu-bound, unvectorized by the JIT \\
\hline
\end{tabular}
\end{table}
\subsection{Measurements}
Measurements are performed using timers inside a caller method. The caller executes the method we want to benchmark with a given number of iterations then returns the mean time spent executing the method. We also iterate over the caller to get the best mean time and then calculate the performance in flop/s performed by the method. Measurements are taken in a steady state considering the warm-up phase to compile the hotpots, but also the memory state. Since the caller invokes the method over the same data set, the memory bandwidth reaches its maximum for the given memory complexity. The caller arithmetic intensity is almost equal to the arithmetic intensity of the method multiplied by the method invocation count performed by the caller. Theoretically, by inlining at source code, we could increase the memory bandwidth and peak performance by swapping the loop iterating over the data with that iterating over the method computation. However, by default compilers do not perform this kind of optimization which can disturb numerical precision by switching floating point operations.
\section{Results and analysis}
\begin{figure}[h]
\centering
\includegraphics[width=\columnwidth]{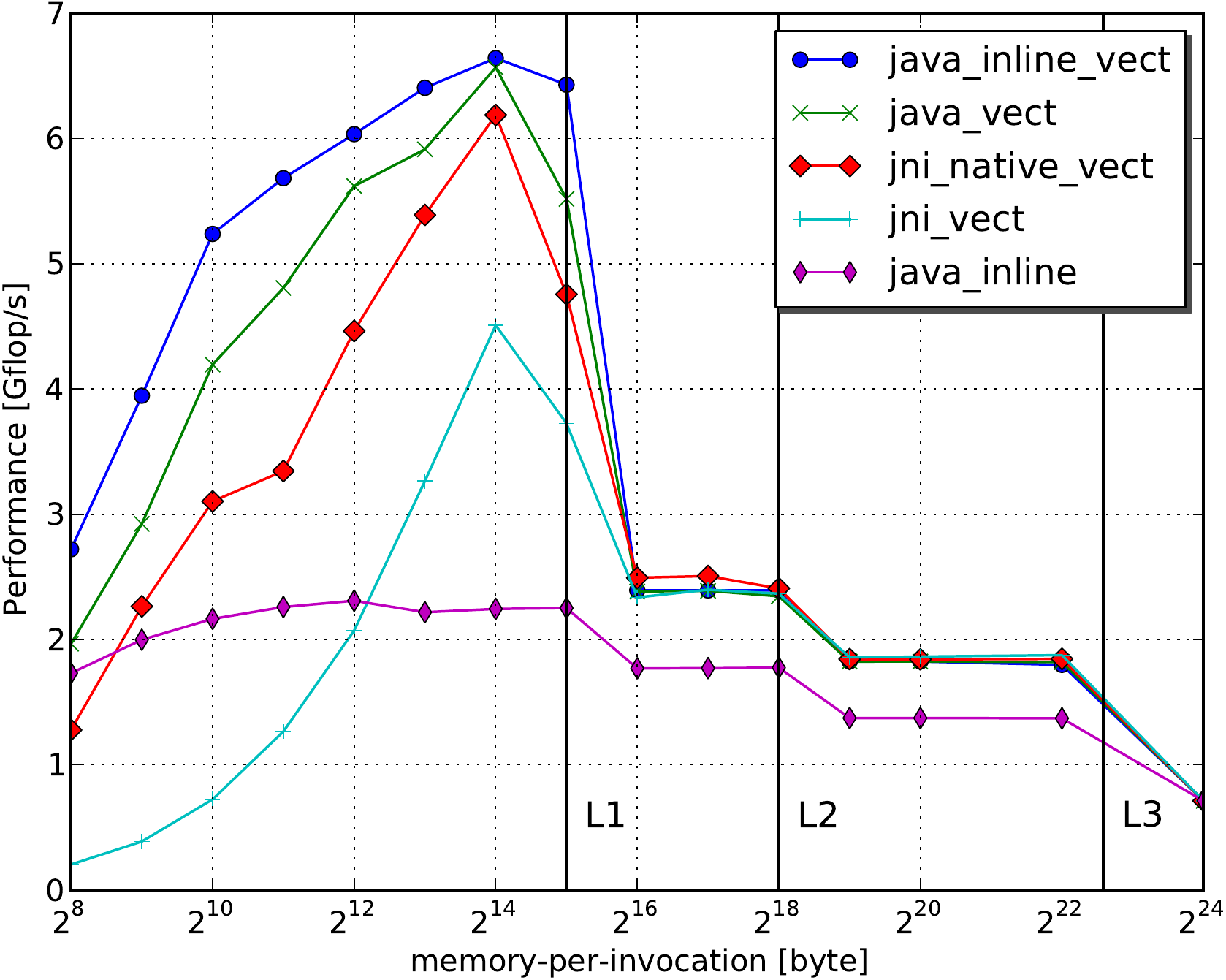}
\caption{Array addition performance profile}
\label{arraySum}
\end{figure}

\begin{figure}[h]
\centering
\includegraphics[width=\columnwidth]{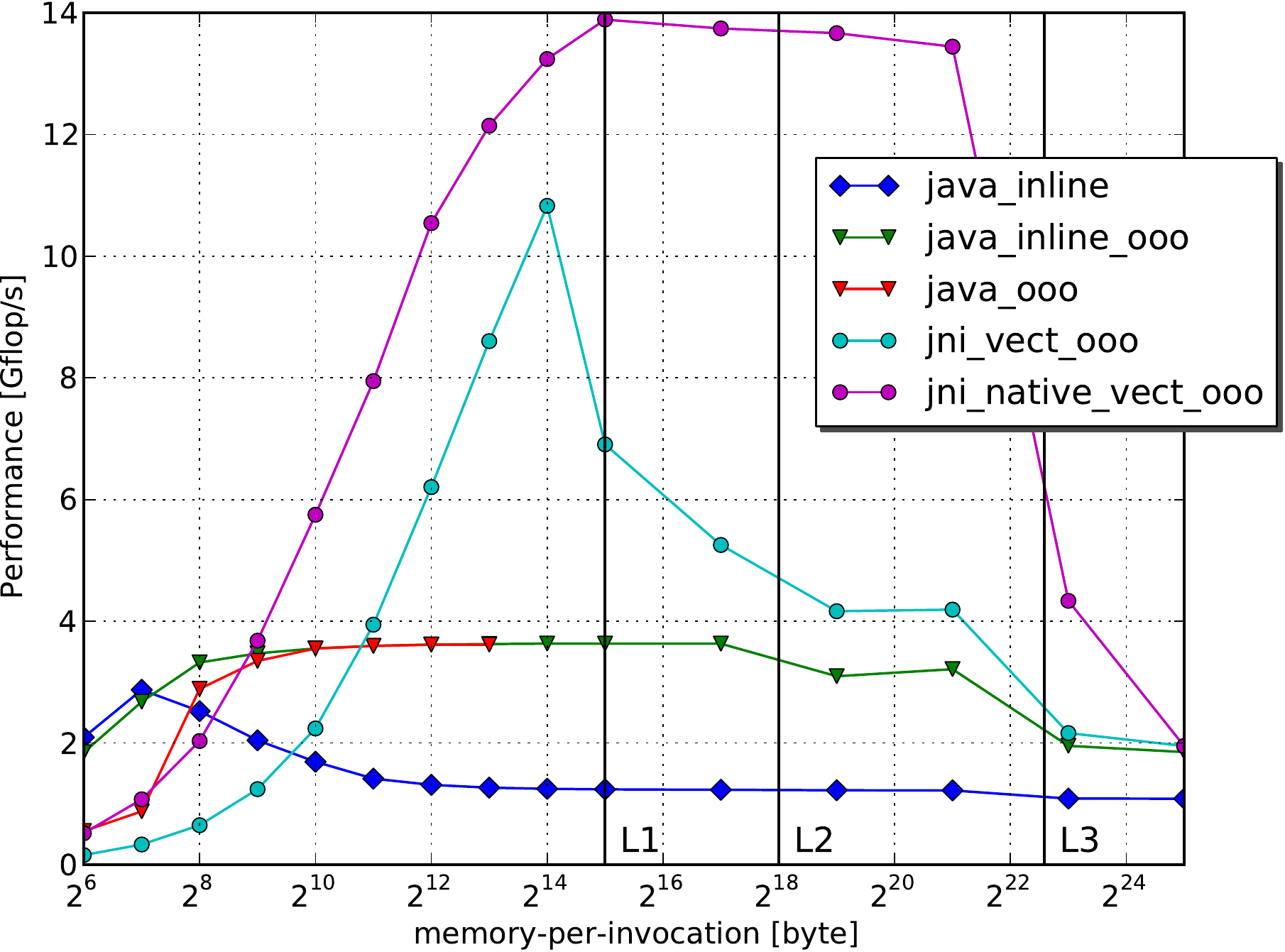}
\caption{Horizontal sum performance profile}
\label{sumReduction}
\end{figure}

\begin{figure}[h]
\centering
\includegraphics[width=\columnwidth]{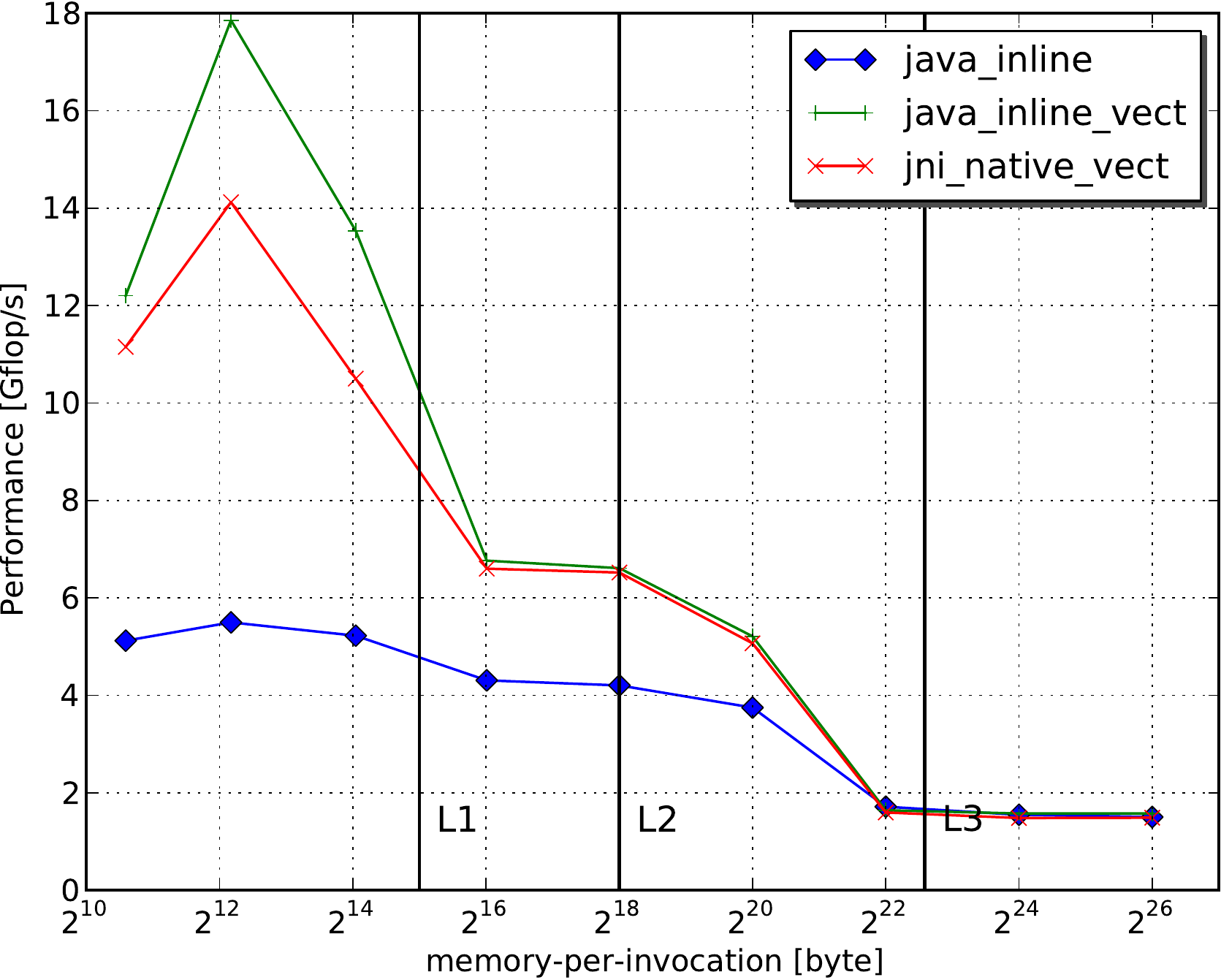}
\caption{Horner coefficient-1st performance profile}
\label{hornerdc1}
\end{figure}

\begin{figure}[h]
\centering
\includegraphics[width=\columnwidth]{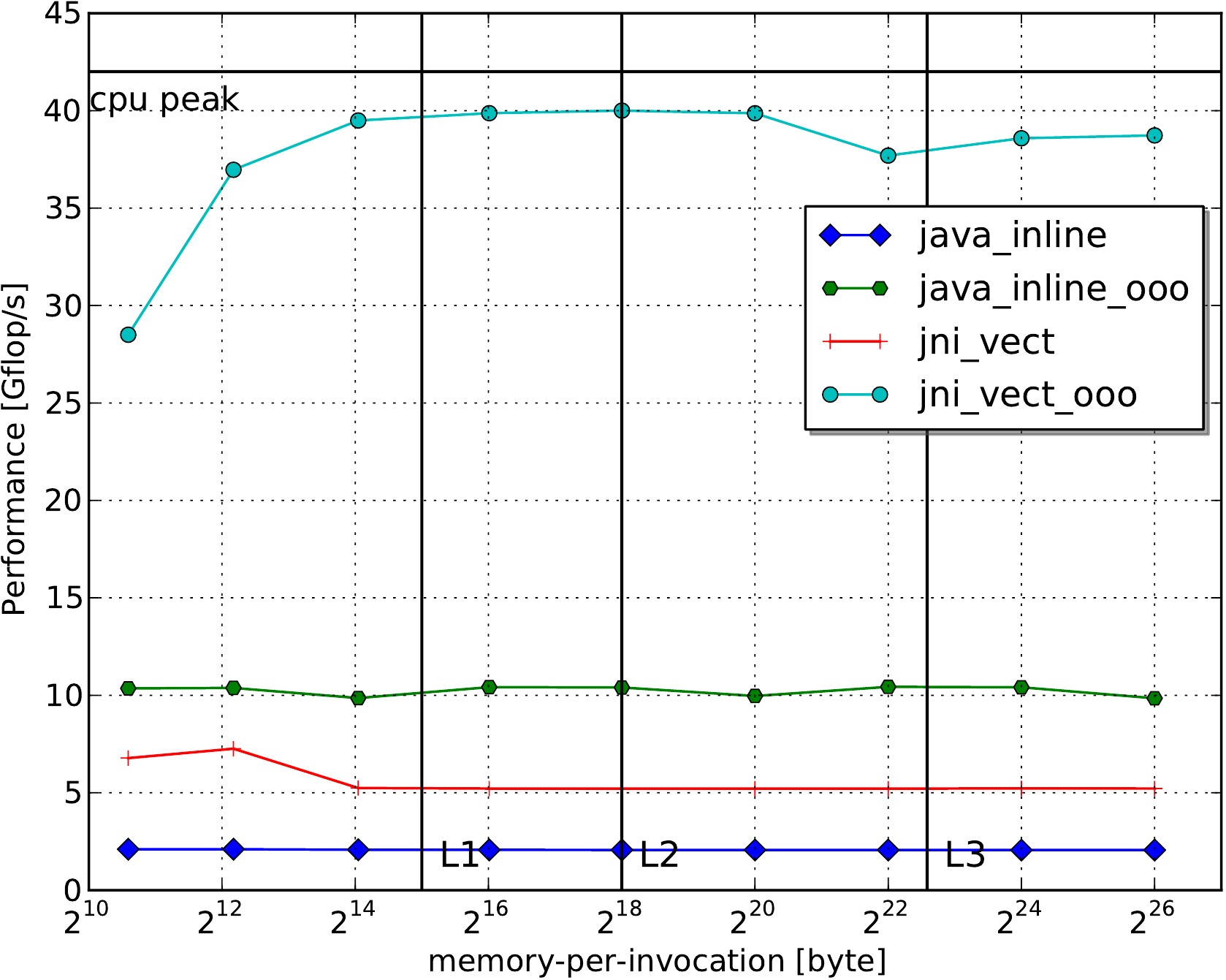}
\caption{Horner data-1st performance profile}
\label{hornerd1}
\end{figure}

For each kernel we provide the performance profile for several different implementations (Figures \ref{arraySum}, \ref{sumReduction}, \ref{hornerdc1} and \ref{hornerd1}). Performance profiles are plotted as a function of the memory-per-invocation. Thereby we can analyze the result by accounting for the cache hierarchy. 

\label{Results}

\subsection{Most efficient implementation}
The performance profile allows to locate the most efficient implementation for a given range of memory-per-invocation. This range is supposed to be known for a particular calling context. \\
The most efficient implementation of the \textit{Array addition} kernel is the vectorized and inlined Java version (\textit{java\_inline\_vect}) for the whole range of memory-per-invocation considered. This is an example where the JIT works as well as GCC, hence there is no benefits from using JNI in such a case.\\
Concerning the \textit{Horizontal sum} kernel, the best implementation depends on the considered range of memory-per-invocation. Since native memory breaks the Java language, we don't consider JNI with native memory  as eligible. Thus, until 2kB the best implementation is the inlined Java version with an out-of-order optimization (\textit{java\_inline\_ooo}). From 2kB the most efficient implementation is the vectorized and out-of-order optimized JNI version (\textit{jni\_vect\_ooo}). This kernel is an example where using static compilation leads to benefits but only when the amount of computation is sufficient to cover the invocation cost. The performance profile allows to quantify this threshold. \\
For the \textit{Horner} kernel we provide two different algorithms. The coefficient-1st iterates over the polynomial coefficients at first then over the input values. The data-1st iterates over the data-values at first then over the coefficient values. The most efficient implementation considering both algorithm is provided by the JNI version which is vectorized and out-of-order optimized (\textit{jni\_vect\_ooo}). This is true for the whole range of memory-per-invocation considered.

\subsection{Vectorization}
The efficiency of vectorization is related to the memory bandwidth. To observe a substantial speed-up, the bottleneck must be computational. For memory-bound kernels, we observe a performance decrease along with the memory-per-invocation that reduces the memory bandwidth.\\
Considering the \textit{Array addition} kernel for a range of memory-per-invocation between L2 size and L3 size, all vectorized implementations are almost equal with a benefit of at least 30\%. For memory-per-invocation greater than L3 size there is no longer a vectorization benefit since the memory bandwidth is the major bottleneck. Vectorization is really significant if the memory-per-invocation is lower than the L1 size and optimal for a memory-per-invocation of 16kB. \\
Java implementations of the \textit{Horizontal sum} kernel are not vectorized since the JIT doesn't support the vectorization of reduction idioms. Since the kernel is memory bound, vectorized JNI implementations are sensible to the memory bandwidth and from a memory-per-invocation greater than the L3 size we have no more vectorization benefits. This threshold is greater for this kernel than the \textit{Array addition} one because its arithmetic intensity is also greater. \\
Contrary to the \textit{array addition} and the \textit{horizontal sum} kernels, the Horner algorithm has an arithmetic intensity resolved at run-time since it depends on the polynomial degree and the number of points to evaluate. For the range of inputs considered, the kernel is CPU-bound (the polynomial degree is fixed to 64 and the number of points greater than 4). However, the coefficient-1st version is affected by the memory bandwidth like a memory-bound kernel. The coefficients fills a constant amount of memory that is relatively small and fits into the L1 data-cache (512 bytes for a 64-degree polynomial). However, for each coefficient this version needs to load all the input data which causes a lot of cache misses when the amount of data exceeds the L1 size. And this drastically reduces the memory bandwidth. The data-1st version iterates over the coefficient array through a dependency chain for each data. Data is loaded only one time and since the coefficient array fits onto the L1 data-cache the memory bandwidth is optimal. However the data-1st version is not vectorized by the JIT because it exposes a reduction idiom. Vectorization of the coefficient first implementation leads to a significant speed-up factor of 2 but only for a memory-per-invocation lower than L3 size. Beyond that size, vectorization has no impact due to the memory bottleneck.
 
\subsection{Out-of-order optimization}
Out-of-order optimizations are performed by dividing a dependency chain into several independent sub dependency-chains. Kernels which expose a dependency chain are the \textit{Horizontal sum} and the \textit{Horner data-1st}. The number of divisions to achieve an optimal out-of-order execution depends on the CPU execution units requested. For example, we need 4 dependency-chains for the \textit{Horizontal sum} and 8 for the \textit{Horner data-1st}. Mixing out-of-order and vectorization leads to an optimal instruction level parallelism and performance. Since the JIT can not vectorize such idioms, JNI proves to be an efficient alternative in such case. Considering the \textit{Horner data-1st kernel}, mixing vectorization and out-of-order
allows performance to reach the CPU peak.

\subsection{Call-bound part}
For all implementations, the performance penalty when the memory-per-invocation decreases is initially due to the constant cost of the loop in the measurements. In this decreasing part, implementations are call-bound since the invocation cost is the main bottleneck. Considering the \textit{Array addition} kernel, the unvectorized Java implementation shows that this cost is rather limited since its performance remains relatively stable. The JNI and Java invocation cost is observed comparing the un-inlined version of the Java method with the JNI method which use native memory. Disabling inlining for Java methods shows that the invocation cost is limited since inlining yields a mean improvement of about 20\%. Finally the comparison between native memory and heap memory for JNI methods shows that the call bound is mainly due to the invocation of JNI callbacks, first, to get a pointer from the Java array inside the heap and then to release it when the work is done. This causes a penalty of about 40\%. The observation is the same for the \textit{Horizontal sum} kernel where using native memory leads to a benefit of 60\%.

\subsection{Data Alignment}
\begin{figure}[h]
\centering
\includegraphics[width=\columnwidth]{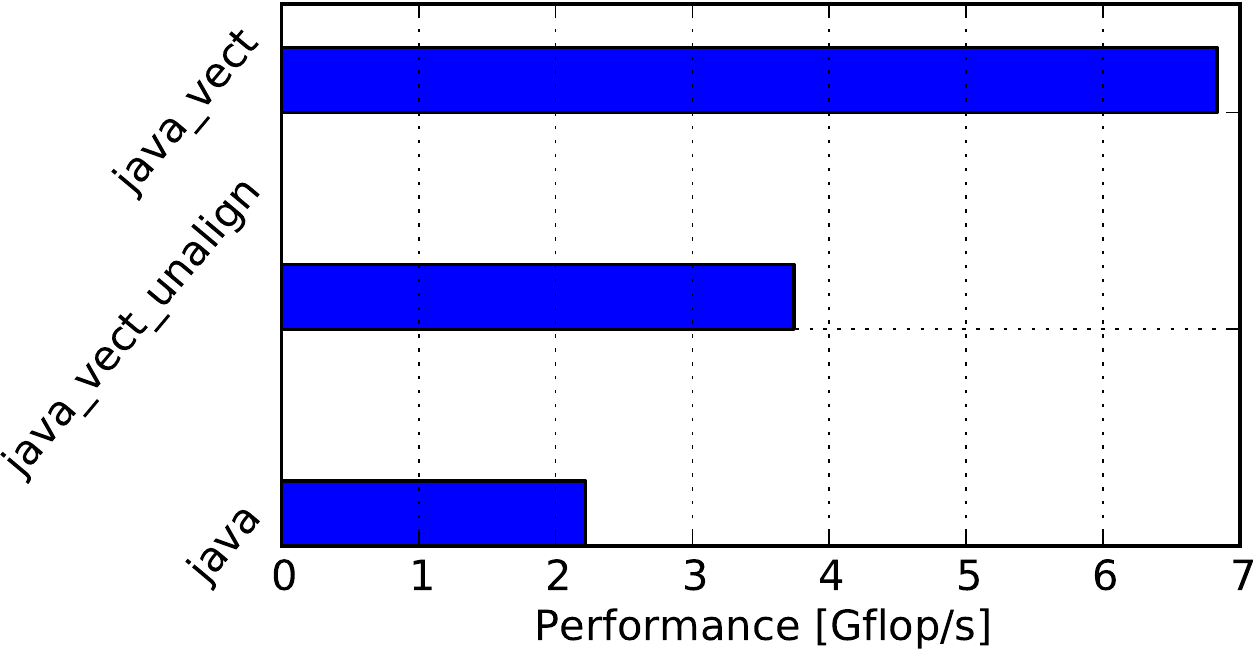}
\caption{\textit{Array addition} performance for a memory-per-invocation equal to 16kB}
\label{alignfig}
\end{figure}

Misalignment performance penalty occurs when four-double packets are not aligned on 16-bytes. Peeling is used by compilers to begin vectorization at the first aligned packet. In Java we can't explicitly align data at allocation, moreover the garbage collector can move the data and as a result change their alignment. In Java double arrays are at least 8-byte aligned. Considering the \textit{Array addition} kernel, if one array is aligned on 8-byte and the second one on 16-byte then peeling is inefficient and memory access will be unaligned for one array. As showed in Figure \ref{alignfig}, this leads to a performance decrease of around 45\% which is totally unpredictable since alignment is resolved at allocation (invisible to Java programmer) and can be modified by the garbage collector.

\section{Related works}
There is so far no performance comparison between pure Java and JNI that takes into account both the overhead of JNI calls and potential deeper optimization provided by the static compilation. Nuzman et al. presented a split auto-vectorization framework \cite{ref_vapor} combining dynamic compilation with an off-line compilation stage aiming at being competitive with static compilation while conserving application portability. Parri et al. \cite{ref_parri} designed an API called jSIMD that uses JNI as a bridge to map Java code to SIMD instructions using vectorized data of various types. Regarding JNI performance issues, Grimmer et al. implemented the Graal Native Function Interface (GNFI) for the Graal Virtual Machine \cite{ref_GNFI} as an alternative to JNI. GNFI aims to mitigate all the disadvantages met using JNI both concerning programming flexibility than performance. Stepanian et al. \cite{ref_inlineJNI} proposed an approach for the IBM TR JIT compiler to widen the compilation span by inlining native code. Finally, Kurzinyec and Sunderam \cite{ref_JNIPerf} studied the performance of different JNI implementations for several different JVM.

\section{Conclusion}
In this paper we have presented a performance analysis 
for a set of micro-kernels considering the JIT vectorization limitation and the JNI invocation cost. By plotting the performance profile for several different implementations of a kernel, we have aimed to select the most efficient implementation for a specific amount of computation. We have showed that one major performance issue in Java concerns reduction kernels that are not vectorized. As a consequence Java implementations may suffer from severe performance penalties compared to JNI ones. By using native memory we showed that JNI suffers from a major performance penalty coming from callbacks use to access data inside the Java heap. 

\bibliographystyle{abbrv}
\bibliography{references}

\end{document}